\documentclass[preprint,showpacs,preprintnumbers,amsmath,amssymb,superscriptaddress]{revtex4}

\include{ams}
\usepackage[english]{babel}
\usepackage{amsmath,amssymb,amsthm}
\usepackage{dcolumn}
\usepackage{bm}
\usepackage{graphicx}

\begin{document}

\title[Anomalous Isotopic Effect of Tunneling States in NbTi-H/D]{Anomalous Isotopic Effect of Tunneling States in NbTi-H/D}

\author{S. Sahling}\email{sahling@physik.tu-dresden.de}

\affiliation{Institut f\"{u}r Festk\"{o}rperphysik, TU Dresden,
01062 Dresden, Germany}

\author{S. Abens}
\affiliation{Institut f\"{u}r Festk\"{o}rperphysik, TU Dresden,
01062 Dresden, Germany}

\author{V.L. Katkov}\email{katkov@theor.jinr.ru}
\affiliation{Bogoliubov Laboratory of Theoretical Physics, Joint
Institute for Nuclear Research, 141980 Dubna, Moscow region, Russia}

\author{V.A. Osipov}\email{osipov@theor.jinr.ru}
\affiliation{Bogoliubov Laboratory of Theoretical Physics, Joint
Institute for Nuclear Research, 141980 Dubna, Moscow region, Russia}

\begin{abstract}
The thermal conductivity, heat capacity and heat release of NbTi,
NbTi-H and NbTi-D were measured at low temperatures. All three
systems reveal low temperature anomalies typical for structural
glasses. It is shown that above a concentration of 2\% H or D the
tunneling states of the NbTi matrix disappear. Therefore,
for higher concentration it is a good system to proof how the change of
the mass of tunneling atoms influences the low temperature
anomalies. For the heat capacity we found the expected isotopic
effect. However, the anomalous isotopic effect observed for the
heat release data cannot be explained within the standard
tunneling model. A surprising result is that tunneling systems
with very high barrier heights, even the systems with the maximum
barrier height, influence remarkable the heat capacity and the heat
release data at low temperatures. As a possible origin,
we consider large-scale fluctuations in thermal expansion
which could generate anomalous two-level systems.

\end{abstract}



\maketitle

\section{Introduction}

All investigated amorphous solids show quite universal low
temperature anomalies of the acoustic, dielectric and thermal
properties. Although most of these anomalies can be
explained within either the phenomenological tunneling or soft
potential model, their microscopic origin is not yet cleared up.
Surprisingly, a glassy-like behavior was also observed in some
crystalline solids at low temperatures~\cite{W1,W2,W3,Cleve,Ray,Chur}.
In some plastically
deformed crystals (Al,Ta, and Nb) the intrinsic properties of linear defects
(dislocations) were found to be of importance~\cite{Chur}. However, the analysis shows
that this picture is not universal. For example it is not appropriate
for the description of observed glass-like anomalies
in $\omega-\beta$ alloys.

One of the physically interesting solids is a stabilized cubic
zirconium dioxide~\cite{Abens}. In experiments with a Ca
stabilized ZrO$_2$ single crystal we found together with a typical
glassy behavior of the low temperature properties (like the
thermal conductivity, heat capacity, ultra sound velocity and
internal friction) a {\it giant heat release}, which exceeds
typical values for amorphous and other glass-like crystalline
solids by roughly two orders of magnitude. The analysis of these
data shows that the giant heat release is caused by tunneling
systems with extremely large barrier heights. An interesting
question is why the relaxation time of these tunneling systems is
so short that they can contribute to the heat release in a time
scale of minutes? Unfortunately, any theoretical attempts to
explain this phenomenon are failed. The reason is that the giant
heat release depends on many factors such as the maximum barrier
height of the distribution, a zero point energy of tunneling
systems, cooling rates, and the thermal expansion of the material.

In this paper, we focus on a study of the low-temperature behavior
of NbTi~\cite{Sahling2}. Ti atoms in  NbTi  act like traps for
atomic hydrogen (H) or deuterium (D).  It seems that at low
temperatures H or D atoms can change their positions near Ti atoms
by a tunneling process thereby causing typical for structural
glasses anomalies in the heat capacity \cite{Neumaier} or in the
long-time heat release \cite{Schwark}. Since the hydrogen in a
given NbTi sample can be easily replaced by deuterium, we are
permitted to investigate the influence of the mass of tunneling
atoms on the low temperature anomalies, which is impossible for
structural glasses. Therefore we have measured the thermal
conductivity, heat capacity and heat release of NbTi with nearly
the same concentration of H or D.

An additional problem is that the internal friction of pure
polycrystalline Nb$_{x}$Ti$_{1-x}$  with 0.2 $<x<$ 0.6 also shows
low temperature anomalies typical for structural glasses
\cite{Cleve}. Actually, our measurements of pure
Nb$_{0.37}$Ti$_{0.63}$ reveal evidence of a glassy behavior for
other main thermal characteristics. Thus, additional tunneling
states are caused by atoms in the matrix. These tunneling systems
are originated from local phase fluctuations between the $\beta$ -
and $\omega$ - phases, which are close in their energy
\cite{Cleve}. However, investigations of the crystal structure of
NbTi show that a small amount of hydrogen leads to a thermally and
mechanically stable single $\beta$ - phase structure \cite{Ilyin}
and we \textit{can assume} that these additional tunneling states
disappear above some critical concentration of H or D.

\section{Model}

Let us start from some useful theoretical background. The
experimental data will be analyzed within the standard tunneling
model (STM)~\cite{Anderson1, Phillips1} which is based on the
assumption that there exist two-level systems (TLSs) with a
constant distribution of their asymmetry energy $\Delta$ and
tunneling parameter $\lambda$
\begin{equation}\label{eq2}
P(\Delta, \lambda) = P = const, \qquad \Delta<\Delta_{max}, \lambda<\lambda_{max},
\end{equation}
where $P$ is a so-called "spectral density". For the thermal conductivity, $\kappa$,
STM yields
\begin{equation}\label{eq1}
\kappa/T^2 = \frac{2 \pi
k_B^3}{3h^2}\frac{\rho}{P}\left(\frac{v_l}{\gamma_l^2} +2
\frac{v_t}{\gamma_t^2}\right)
\end{equation}
where $\rho$ is the mass density, $\gamma_{l,t}$ are the coupling constants between
TLSs and phonons, and $v_{l,t}$ are the sound velocities of the
longitudinal and transversal sound waves. As is seen, the
thermal conductivity is very sensitive to magnitudes of
the coupling constants.

On the contrary, the specific heat depends quite weakly on the coupling
constants. This fact allows one to extract $P_C$ from experiment (here and
below we use notation $P_C$ for the parameter $P$ when it was extracted
from the heat capacity measurements). Explicitly,
\begin{equation}\label{eq3}
 C_{TLS} =  \frac{\pi^2k_B^2}{12} P_C T \ln\frac{4 t_0}{\tau_{min}},
\end{equation}
where  $t_0$ is a characteristic time of the heat capacity
measurement and $\tau_{min}$ is the minimum relaxation time.
$\tau_{min}$ can be estimated from the general expression for the relaxation
time $\tau$ of a tunneling process
\begin{equation}\label{eq4}
\tau  = \left[A (E \Delta_0^2/k_B^3) \coth(E/2 k_BT)\right]^{-1}
\end{equation}
with $E = \sqrt{\Delta_0^2 + \Delta^2}$ and
\begin{equation}\label{eq5}
A = \frac{8\pi^3k_B^3}{\rho h^4} \left(\frac{\gamma_{l}^2}{v_l^5}
+ 2\frac{\gamma_t^2}{v_t^5}\right)
\end{equation}
on conditions that $E=\Delta_0= 2k_B T$.

The long-time heat release is of our main interest here because it
is most sensitive to the expected isotopic effect. After a rapid
cooling of a sample from some initial (equilibrium) temperature
$T_{1}$ to a final (phonon) temperature $T_{0}$, the tunneling
systems with relaxation time $\tau$ longer than the time necessary
for cooling remain in a non-equilibrium state. Their further
relaxation to a new equilibrium state leads to an energy transfer
from tunneling systems to phonons~\cite{Phillips2}
\begin{equation}\label{eq6}
dQ/dt\equiv\dot{Q}\left(T_{1},T_{0},t\right) = \frac{\pi^2 k_B^2}{24} P_Q V^{s}
(T_1^2 - T_0^2) t^{-1}  ~~\textrm{for}~~T < T^*,
\end{equation}
and
\begin{equation}\label{eq7}
\dot{Q}_{max} = \frac{\pi^2 k_B^2}{24} P_Q V^{s} (T^{*2} - T_0^2)
t^{-1}
 ~~\textrm{for}~~T > T^*,
\end{equation}
where $P_Q$ is the tunneling parameter extracted from the heat release
measurements, $V^{s}$ is the volume of the sample, and the freezing
temperature $T^*$ is given by~\cite{Parshin1}
\begin{equation}\label{eq8}
T^* = \frac{V_{eff}}{\ln (k_BT^{*2}/\tau_0|R^*|V_{eff})}.
\end{equation}
Here  $V_{eff}$ is the effective barrier height of TLS causing
the heat release at given time $t_0$, $R^*$ is the cooling rate at
the freezing temperature $T^*$, and $\tau_0$ is the constant of the
thermal activation rate. The saturation of the heat release above
$T^*$  is a consequence of the thermal activation process which
dominates at high enough temperatures. The relaxation time at
higher temperatures becomes so short that TLSs  with given
$V_{eff}$ and higher energies  $\Delta \approx E > 2 k_BT^*$ reach
the equilibrium during the cooling process and do not contribute
to the heat release.

Obviously, within the STM $P = P_C = P_Q$. Numerical calculations
show that 95\% of the heat release observed at fixed time  $t_0$
is produced by tunneling systems whose relaxation times lie in the
range of $0.2 t_0 < \tau < 20 t_0$. These TLSs are located in a
quite small region of the distribution function $P(\lambda)$ with
$\Delta\lambda = 2.6$ for a typical time of the heat release
experiments $t_0 = 1$h and $\lambda$ varying between 15 and 20.
For instance, in structural glasses this range amounts to about
20\%  of the effective tunneling parameter $\lambda_0$ determined
by the equation $\tau = t_0$. Thus, all TLSs causing the heat
release at time $t_0$ are contained in a small range of
$P(\lambda)$ around $\lambda_{0}$. If we replace H by D and
measure the heat release at the same time $t_0$ we observe the
heat release of TLSs with equal $\lambda_{0}$. The relation
between $\lambda_0$ and $V_{eff}$ is given by \cite{Phillips1}
\begin{equation}\label{eq9}
\lambda_0 = \frac{d}{\hbar}\sqrt{2m V_{eff}},
\end{equation}
where $d$ is a distance between potential minima and $m$ is the
mass of tunneling atoms. Therefore, if we replace H by D and
assume that the distribution of TLSs is completely determined by
NbTi matrix, the heat release at the same time $t_0$ will be
caused by TLSs with markedly smaller barrier heights. Indeed, at
equal $\lambda$ one obtains $V_{eff}^H/V_{eff}^D = 2$ as long as
$m^H/m^D = 1/2$. According to Eq.~(\ref{eq8}) this results in a
large isotopic effect in the freezing temperatures
\begin{equation}\label{eq10}
\frac{T^{*H}}{T^{*D}} \cong \frac{V_{eff}^H}{V_{eff}^D} = 2.
\end{equation}

In addition, the distribution parameter should also be changed.
From Eq.~(\ref{eq9}) follows
\begin{equation}\label{eq11}
\frac{P^H}{P^D} = \frac{\lambda_{max}^D}{\lambda_{max}^H} =
\sqrt{\frac{m^D}{m^H}}\approx 1.4,
\end{equation}
since the places of D and H and the barrier heights between them
are determined by the lattice of NbTi (i.e. $d^H=d^D, V_{max}^H = V_{max}^D)$
and we expect for equal concentrations the same number of TLSs ($N^H = N^D$),
which is proportional to $P\lambda_{max}$.

Thus we expect a
drastic change in the maximum value of the heat release
\begin{equation}\label{eq12}
\frac{\dot{Q}_{max}^H}{\dot{Q}_{max}^D}= \frac{P^H}{P^D}
\left[\frac{(T^{*H})^2-T_0^2}{(T^{*D})^2-T_0^2}\right]\approx\left(\frac{m^D}{m^H}\right)^{5/2}\approx
5.7
\end{equation}
for  $T_0 \ll T^*$. In other words, the heat release caused by D
saturates at much lower temperature  $T^{*D} =  T^{*H}/2$ with an essentially
smaller peak value.

Assuming that the double-well potential can be formed by two
identical harmonic potentials with a shifted position of their
potential minima, the tunneling parameter is written as~\cite{Tielburger}
\begin{equation}
\lambda_0 = \frac{V_{eff}}{E_0},
\end{equation}
where E$_{0}$ is the zero-point energy in a single harmonic
well. One obtains
\begin{equation}
\frac{E_0^{H}}{E_0^{D}} = 2.
\end{equation}
This relation will also be tested in our experiments.

\section{Experimental}

Three samples were prepared from the same Nb$_{0.37}$Ti$_{0.63}$
target: a large one ($V = 10$ cm$^3$, $m = 60.17$ g) for the heat
release experiments, a small one $(0.86$ g) for the heat capacity
and a long one for the thermal conductivity measurements. Firstly,
all these samples were heated up to 650$^\circ$C  for 24 h at a
pressure of $p = 10^{-6}$ mbar to reduce the concentration of
trapped H, N and O to values less than 10 at ppm. After experiments
with pure NbTi the samples were charged with H by 650${}^\circ$C and
different pressure of hydrogen gas. The H-concentration in the
samples was calculated from the mass difference and is given by the
ratio of H atoms to the total number of Ni and Ti atoms. In this way
H-concentrations ranging between 1 and 50 \% were obtained. After
experiments with NbTi-H  the hydrogen was again reduced to zero (in
fact the initial mass of the pure NbTi was obtained again) and than
charged with D in the same way.

Heat capacity and thermal conductivity experiments were performed in
a ${}^3$He -${}^4$He - dilution refrigerator, the heat release was
measured mainly at $T_0 = 1.34$ K after cooling from different $T_1$
in a  ${}^4$He cryostat. For the heat capacity measurements both the
pulse technique and the relaxation time method were used. A special
cryostat with a low and very stable parasitic heat flow to the
sample  $\dot{Q}_{par} = 1.5$ nW was used. The rapid cooling of the
sample was realized by a mechanical heat switch. The heat release of
the sample was determined from the temperature drift of the sample
$\dot{T}$
\begin{equation}
 \dot{Q} = (C_p +C_{ad})\dot{T} + (T(t) - T_0)/R_{hl} -
 \dot{Q}_{par},
\end{equation}
where $C_{ad}$ is an addendum heat capacity (the thermometer, the
heater, the plate). The total heat capacity $C_p + C_{ad}$ (the time
dependence of $C_p$ can be neglected) and the heat link $R_{hl}$
between the sample and the sample chamber were measured separately.
The sample was fixed by thin nylon threads, which causes by open
heat switch together with the electrical wire of the heater and the
thermometer the heat link $R_{hl}$.  In some experiments, where the
temperature drift was very strong, the heat switch was some times
closed during the measurement to reduce the sample temperature
$T(t)$ to $T_0$ again. Thus, $T(t)$ was close to $T_0$ during the
whole time of the heat release measurement $(T(t) - T_0 < 10$ mK at
$T_0 = 1.34$ K).

\section{Results and Discussion}
\subsection{Thermal conductivity}

The thermal conductivity of pure NbTi is shown in Fig.~1 (black
squares).
\begin{figure} \centering
\includegraphics[width=8cm,  angle=-90]{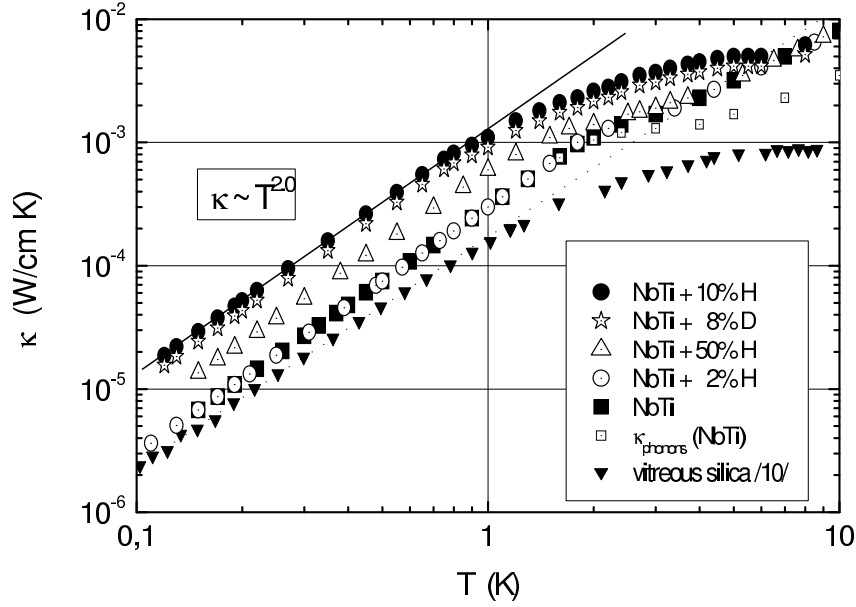}
\caption{The thermal conductivity of NbTi, NbTi-H and NbTi-D for
different concentrations  of H and D. Below 1 K the thermal
conductivity is strongly proportional to  $T^2$ (solid line).}
\end{figure}
Below  1 K the heat capacity is exactly proportional to $T^2$ with
an absolute value close to that of vitreous silica (black
triangles). Thus, the thermal conductivity of polycrystalline NbTi
exhibits low temperature anomalies typical for structural glasses
in accordance with glassy behavior of the internal friction,
observed for the same material in \cite{Cleve}. When we increase H
or D concentrations in NbTi the power law remains always
unchanged. At the same time, we observed a surprising dependence
of the absolute value. Namely, the thermal conductivity of NbTi
with 2 \% H shows the expected result being about  6 \% smaller
the value of pure NbTi, i.e. one has an additional scattering on
the tunneling states produced by the hydrogen atoms. However, the
thermal conductivity of NbTi with  10 \%  H is 3.4  times larger
the value of pure NbTi. A similar result we get with  D. A simple
explanation of this unexpected behavior is that the local
fluctuations, which are responsible for the tunneling states and
low thermal conductivity of pure NbTi, disappear at higher
concentrations of H or D ($x > 8 \%$). In fact, above some
critical concentration $x_c$ (2\%$<$ $x_c$ $<$ 8\%) the hydrogen
(or deuterium) stabilizes $\beta$ - phases and, as a consequence,
one expects that the glassy behavior at low temperatures
disappears \cite{Ilyin}. At the same time, H (or D) produces new
tunneling states. For this reason, the thermal conductivity is
still glassy-like: it is proportional to $T^2$ but with a much
larger absolute value due to smaller coupling constants. We can
calculate the parameter $P\gamma_t^2$ from Eq.~(\ref{eq1}) with
$v_l = 4360$ m/s, $v_t = 2640$ m/s, $\rho = 6$ g/cm$^3$ and
$(\gamma_l/\gamma_t)^2 = 2.5$ \cite{Parshin1}. The results are
presented in Fig.~2.
\begin{figure} \centering
\includegraphics[width=8cm,  angle=-90]{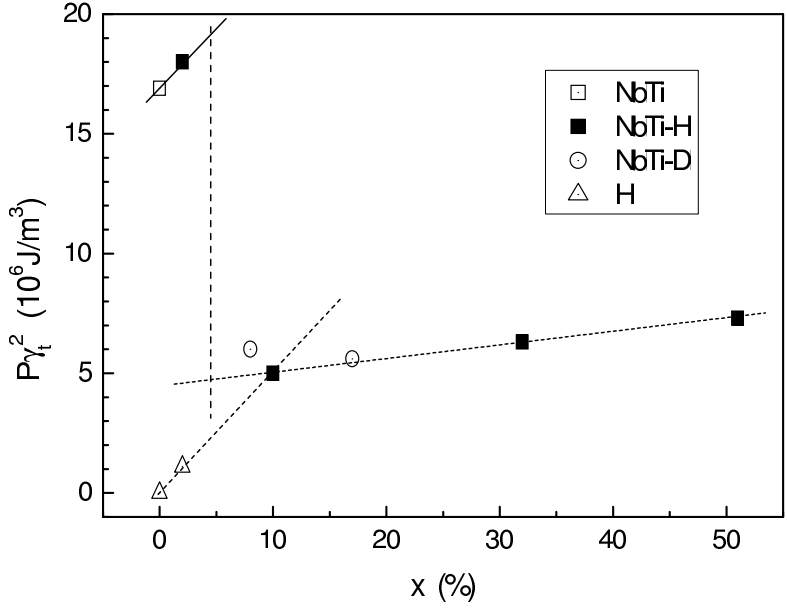}
\caption{The parameter $P\gamma_t^2$ of the standard tunneling
model determined from the thermal conductivity data with
Eq.~(\ref{eq1}).}
\end{figure}
We see clearly the jump of this parameter between  2 and 8 \% H or
D. The number of the tunneling states produced by H or D increases
rapidly with the concentration for  $x < 10\%$  and slowly for
higher $x$.  Since H (or D) is trapped by Ti atoms and changes the
positions around them, it is necessary that the new positions are
free. For higher concentrations this will be less and less the
case and we can even expect a maximum in the dependence of the
number of tunneling states on the H or D concentration. Up to 50\%
this maximum was not observed. Partially this could be masked by
an additional scattering on dislocations, which can appear at high
concentrations of H or D like in case of hydrogen in
cooper~\cite{wampler,kolac}. We did not try to introduce more than
50\% H, since a too high concentration of H destroys the NbTi
sample.

\subsection{Heat capacity}

Typical results of the heat capacity measurements are shown in
Fig.~3.
\begin{figure} \centering
\includegraphics[width=8cm,  angle=-90]{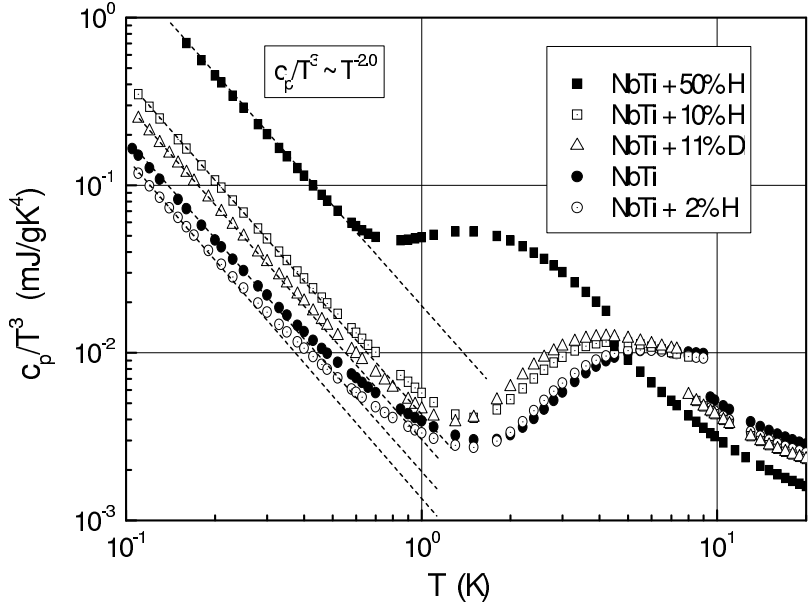}
\caption{The heat capacity of NbTi, NbTi-H and NbTi-D for
different concentration H or D. Below  1 K the heat capacity of
all investigated systems is proportional to $T$ (dashed lines).}
\end{figure}
Below 1 K, where the contribution of the electron systems can be
neglected, the heat capacity is proportional to  $T$ (with a
coefficient denoted here as $a_{ts}$) . This linear term is
roughly proportional to the concentration of H or D: $a_{ts} =
0.37$ mJ/gK$^2$  for $x$ $>$ 2\%. For smaller $x$, $a_{ts}$
increases due to additional tunneling systems caused by the
$\beta$-$\omega$ - fluctuations in NbTi. The distribution
parameter $P_C$ of TLSs can be calculated with
Eqs.~(\ref{eq3})-(\ref{eq5}), $t_0 = 10$ s. Together with
$P\gamma_t^2$ from the thermal conductivity we get also
$\gamma_t$. The most important parameters deduced from the heat
capacity are given in Table I.
\begin{table}[th]\caption{~}
\begin{tabular}{||c||c|c|c|c|c||}
\hline Parameter &  NbTi   &   NbTiH$_{2.5\%}$  & NbTiH$_{10\%}$ &
NbTiD$_{11\%}$ & NbTiH$_{50\%}$\\
\hline\hline
$\Theta_D$  (K)               &243          &253          &280          & 253          &303\\
a$_{ts}$  (mJ/gK$^2$)         &2.0          &1.3          &4.2          & 3.1          & 18\\
$\rho$  (g/cm$^3$)               &6.02         &6.02         &6.04         &6.05          &6.14\\
$v_t$  (km/s)                 &2.64         &2.75         &3.04         &2.75          &3.29\\
$A$ (1/sK$^3$)                &3.9 10$^6$   &3.4 10$^5$   &2.1 10$^5$   & 4.8 10$^5$   &1.4 10$^5$\\
$\tau_{min}$(0.2 K) ($\mu$s)  &4            &46           &76           &33            &11\\
$P_C$ (1/Jm$^3$)              &4.3 10$^{45}$&3.9 10$^{45}$&1.2 10$^{46}$&8.6 10$^{45}$ &4.4 10$^{46}$\\
$P\gamma_t^2$ (J/m$^3)$       &1.7 10$^7$   &1.5 10$^6$   &5.1 10$^6$   &5.5 10$^6$    & ~\\
$\gamma_t$  (eV)              &0.39         &0.12         &0.13         &0.16          &~\\
\hline
\end{tabular}
\end{table}
Since the heat capacity and the thermal conductivity were not
measured exactly at the same concentration, we used the
corresponding extrapolated values $P\gamma_t^2$  from Fig.~2. We
do not calculate the coupling parameter  $\gamma_t$ for $x = 50
\%$ H since a different $x$-dependence of $P\gamma_t^2$ and  $P_C$
was observed for $x$ $>$ 10 \%: while $P\gamma_t^2$ is nearly a
constant for higher $x$, the linear $x$-dependence of $P_C$
remains unchanged up to 50\%. In a case of H we found a different
$x$-dependence for both long (heat capacity) and short (thermal
conductivity) relaxation time. This different behavior can be
explained by making an assumption that the longer relaxation time
(with the correspondingly larger tunneling parameter) is caused by
a longer distance between two positions (see Eq.~(\ref{eq9})).
Many more equivalent positions exist for the long-distance
tunneling in comparison with the short-distance one and,
therefore, a possible maximum of the tunneling rate will be
observed at higher concentrations of H or D.

NbTiH$_{2.5\%}$ contains nearly the same number of TLSs $(P_C)$ as
in a case of pure NbTi.  However, there appear new markedly
different tunneling states. The hydrogen  produces additional TLSs
with an essentially smaller coupling constant  $\gamma_t$  (see
Table I). Thus, 2.5\% H  is enough to stabilize the $\beta$-phase in
NbTi.

\subsection{Heat release}

\subsubsection{Heat release in NbTi}

Fig.~4 shows the measured heat release as a function of time after
cooling starting with different initial temperatures $T_1$.
\begin{figure} \centering
\includegraphics[width=8cm,  angle=-90]{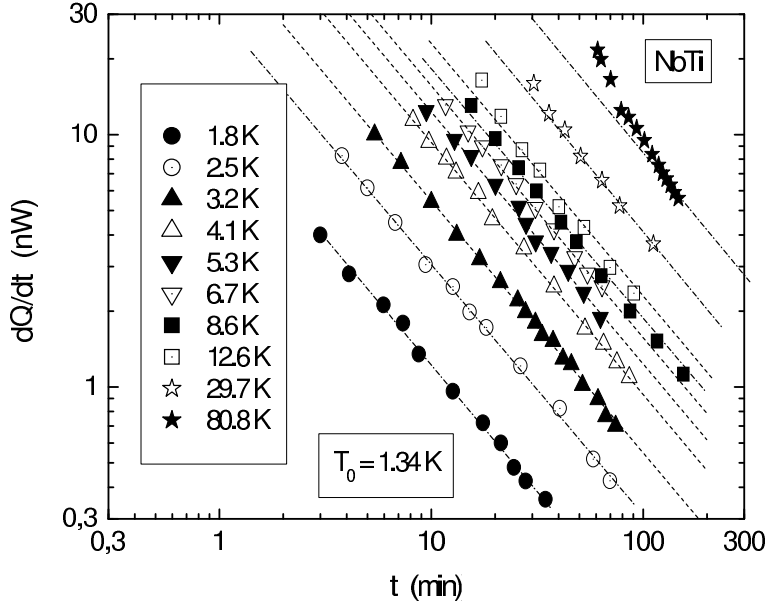}
\caption{The heat release in 10 cm$^3$  NbTi  after cooling from
different initial temperatures  $T_1$  to the final phonon
temperature  $T_0$ as a function of time ($t = 0$ at the beginning of
cooling). The heat release is strongly proportional to $t^{-1}$
(dashed lines).}
\end{figure}
The final temperature  $T_0$ was always 1.34  K. The heat release
is strongly proportional to  $1/t$  (except for $T_1 = 80.8$ K) in
agreement with Eq.~(\ref{eq6}) and thereby with the behavior of
structural glasses. At temperatures $T_1 < 4$ K the heat release
for a fixed time ($t_0 = 30$  min)  is proportional to
$(T_1^2-T_0^2)$ (see the insert in Fig.~5).
\begin{figure} \centering
\includegraphics[width=8cm,  angle=-90]{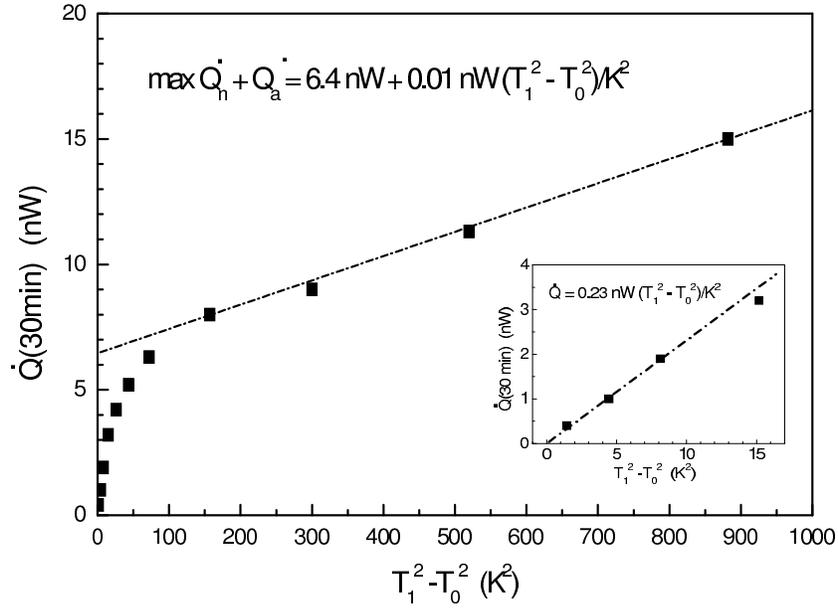}
\caption{The heat release in 10 cm$^3$ NbTi at the fixed time $t_0
= 30$ min as a function of $T_1^2 - T_0^2$ $(T_0 = 1.34$ K).
A linear behavior is found for $T_1 < 4$  K (see the insert) and
for $T_1 > 12$ K. This can be explained by the
assumption that two kinds of TLSs with different freezing
temperatures contribute to the heat release at given time.}
\end{figure}
This allows us to obtain the last free parameter in Eq.~(\ref{eq6}):
$P_Q = 5.3~10^{44}$ J$^{-1}$m$^{-3}$. For higher temperatures  ($T_1
> 4$ K) one could expect the saturation of the heat release as a
function of  $T_1^2-T_0^2$. However, we have obtained a new linear
dependence (see Fig.~5) which is saturated at about 50 K only.
Such behavior is also well known from different structural glasses
\cite{Parshin2}: at given time $t_0$ two kinds of tunneling
systems with different freezing temperatures contribute to the
heat release. For NbTi these two contributions can be easily
separated since two freezing temperatures are very different. At
higher temperatures, the linear dependence  is caused by TLSs with
higher freezing temperature $T^*_a = 47$ K, where the constant at
$T_1^2-T_0^2 = 0$ corresponds to the maximum value of the heat
release of TLSs with the lower freezing temperature $T^*_n = 5.4$
K. These two contributions and the measured heat release as a
function of $T_1^2-T_0^2$ are shown in Fig.~6.
\begin{figure} \centering
\includegraphics[width=8cm,  angle=-90]{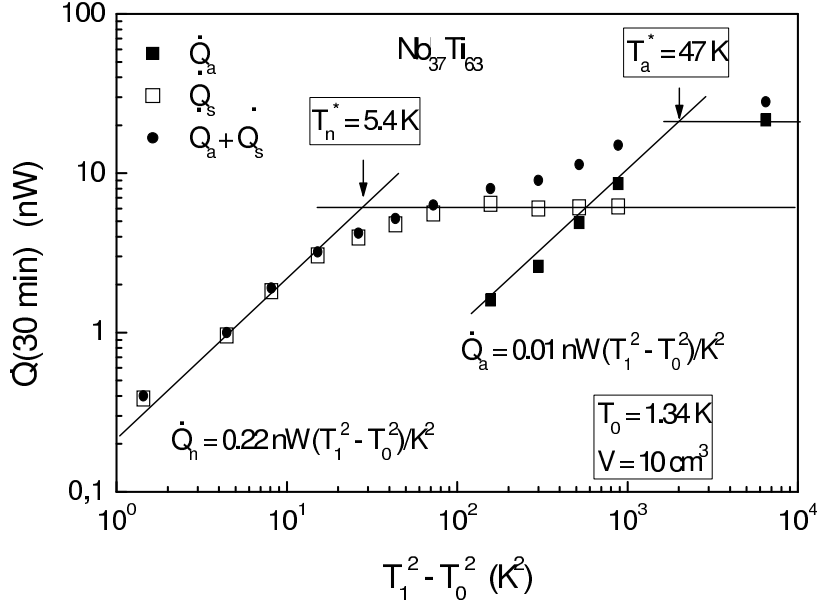}
\caption{The separated contributions of two kinds of TLSs to
the heat release in NbTi at $t_0 = 30$ min.  The white and black
squares show the contribution of TLSs with the freezing temperatures
$T^*_n = 5.4$ K and  $T^*_a = 47$ K, correspondingly. The sum of
these two contributions yields the measured heat release (black
circles).}
\end{figure}
Two corresponding distribution parameters are $P_n = 5.1~10^{44}$
J$^{-1}$m$^{-3}$ and $P_a = 2.3~10^{43}$ J$^{-1}$m$^{-3}$. We will
call TLSs with the lower freezing temperature $T^*_n$ the normal
ones and with the higher freezing temperature $T^*_a$ the anomalous
ones.

The value of $P_a$ is very small and takes only 4\%  of the total
value of $P_Q = P_a + P_n$. However, due to very high freezing
temperature ($T^*_a = 47$ K) their maximum contribution to the heat
release is 3.4 times larger the maximum contribution of the normal
TLSs. The heat release is very sensitive to the presence of TLSs
with a high freezing temperature.

By using of Eq.~(\ref{eq8}) one can calculate now the barrier
heights of TLSs contributing to the heat release. Since the exact
value of $\tau_0$  for NbTi and NbTi-H/D is unknown, we took for
estimations the value of vitreous silica  $\tau_0 = 2\times
10^{-13}$ s \cite{Sahling1}. Further, for TLSs with $\tau = t_0$
we can estimate the corresponding tunneling energy $\Delta_0$ by
using of Eq.~(\ref{eq4})
\begin{equation}\label{eq14}
\Delta_0 \cong 1/\sqrt{A (T^*+T_0) t_0},
\end{equation}
where $T^* + T_0$ is an average energy of TLSs causing the heat
release after cooling from some $T_1  > T^*$. Finally, one can find
from the relation between  $\Delta_0$  and $\lambda$
\cite{Tielburger}
\begin{equation}
\Delta_0 = (2E_0/\pi) [(1 + \lambda)^{1/2} + \lambda^{1/2}]
\exp\left(-\sqrt{\lambda^2 + \lambda}\right)
\end{equation}
the corresponding tunneling parameter $\lambda$ and  the zero-point
energy $E_0$. All the calculated parameters  $V^{eff}, \Delta_{0},
\lambda,$ and $E_{0}$ for the normal (n) and anomalous (a) tunneling
systems are given in Table II.
\begin{table}[th]\caption{~}
\begin{tabular}{||c||c|c|c|c||}
\hline
 Parameter&  NbTi &  NbTiH$_{9\%}$&  NbTiD$_{9\%}$& a-SiO$_{2}$\\
\hline\hline
$P_Q$ (10$^{44}$J$^{-1}$m$^{-3}$)   & 5.3 & 45 &  84 & 2.0\\
$P_{Qn}$ (10$^{44}$J$^{-1}$m$^{-3}$) & 5.1 & 28 &  18.4 & 1.7\\
$P_{Qa}$ (10$^{44}$J$^{-1}$m$^{-3}$) & 0.23 & 17 &  65.6 & 0.3\\
$T^*_n$ (K) &  5.4 & 5.9 &  3.1 & 4.8\\
$T^*_a$ (K) &  47 & 32  & 51 & 14.5\\
$V_n^{eff}/k_B$ (K) &  173 & 193 &  98 & 155\\
$V_a^{eff}/k_B$ (K) &  1606 & 1081 &  1746 & 515\\
$\Delta_{0n}/k_B$ (mK) & 4.5 & 19.2 &  16.2&~\\
$\Delta_{0a}/k_B$ (mK) & 1.7 & 8.9  & 4.7&~\\
$\lambda_n$   & 15.9 & 14.6 &  14.1 & 16.3\\
$\lambda_a$  &  19.0 & 17.0  & 18.1 & 17.2\\
$E_{0n}/k_B$  & 11.0 & 13.3 &  6.9 & 9.0\\
$E_{0a}/k_B$  & 87.3&  63.6 &  96.5 & 25.0\\
\hline
\end{tabular}
\end{table}
Notice that for $\lambda_n$ and $\lambda_a$ we have obtained nearly
equal values. A small difference is caused by a higher average
energy of the anomalous TLSs in comparison with the normal ones.
Within the ordinary STM, the fact that TLSs with very different
barrier heights contribute to the heat release at given time
$\tau_0$ could only be explained  by suggesting the existence of two
kinds of TLSs with very different values of $E_{0n}$ and $E_{0a}$.
However, it will be shown below (see Table III) that the needed
relation for the zero-point energy is strongly distorted for
anomalous TLSs.

There is a different possibility to explain this surprising result.
Namely, since one has only to fulfil the condition $\lambda_a\sim
\lambda_n$ we could expect some kind of instability of the effective
barrier heights of anomalous TLSs. This means that at the freezing
temperature $T^*_a$ the non-equilibrium state of TLSs with the
corresponding barrier height $V^{eff}_{a}$ is frozen in, but after
cooling this effective barrier height is reduced to $V^{eff}_{n}$.
It will be shown below that a strong reduction of the barrier
heights occurs as a consequence of local mechanical stresses during
the cooling processes. In this case, all TLSs should have the same
value of $E_{0n}$.

Another interesting result follows from the comparison of the
distribution parameters deduced from the heat capacity and the heat
release. From the STM we expect that $P_C/P_Q = 1$. However, our
experiment yields $P_C/P_Q = 8$ (see Tables I and II), i.e. there is
an additional contribution to the heat capacity which does not
influence the heat release. Since this discrepancy was also observed
in other materials~\cite{Parshin2, Sahling1, Sahling2} including
structural glasses, we will discuss this problem in more detail in
chapter \ref{435}.

\subsubsection{Heat release in NbTi-H/D}

Figs.~7 and 8 show the heat release of NbTiH$_{9\%}$ and
NbTiD$_{9\%}$ as a function of time after cooling from different
initial temperatures $T_1$ to $T_0 = 1.34 K$.
\begin{figure} \centering
\includegraphics[width=8cm,  angle=-90]{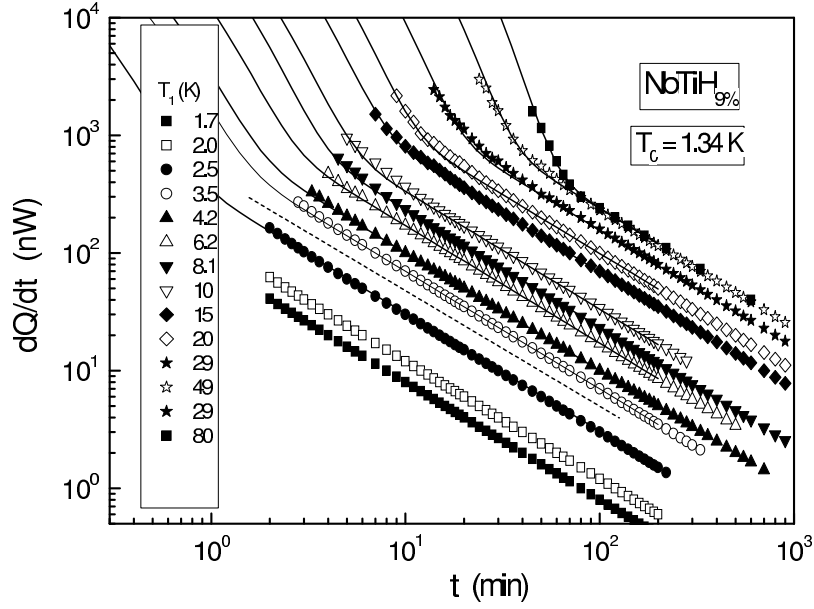}
\caption{The heat release in 10 cm$^3$  NbTiH$_{9\%}$ after
cooling from different initial temperatures  $T_1$  to the final
phonon temperature  $T_0$  as a function of time ($t = 0$ at the
beginning of cooling). The heat release is proportional to
$t^{-1}$ (broken line) except for  $T_1 > 8$ K, where at short
time an additional contribution exists, which relaxes
exponentially.}
\end{figure}
\begin{figure} \centering
\includegraphics[width=8cm,  angle=-90]{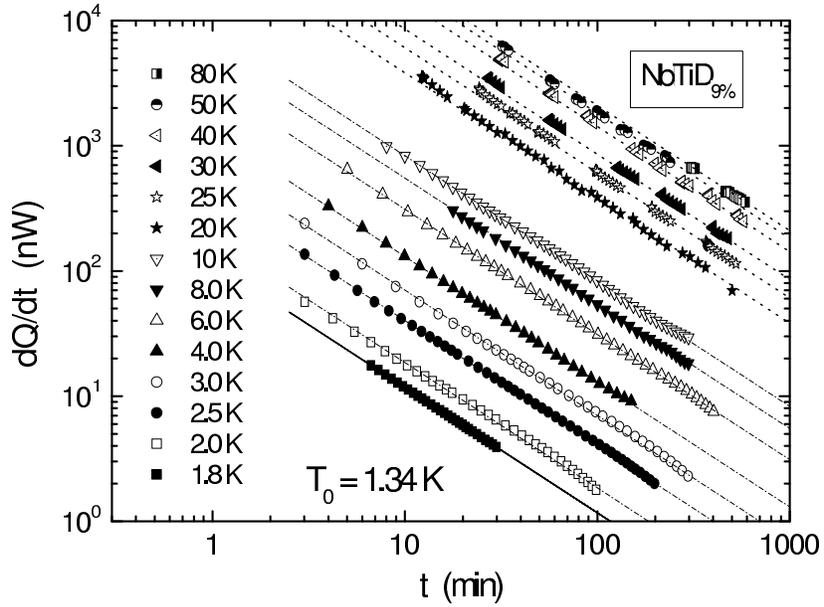}
\caption{The heat release in 10 cm$^3$  NbTiD$_{9\%}$ after
cooling from different initial temperatures  $T_1$  to the final
phonon temperature $T_0$  as a function of time (t = 0 at the beginning
of cooling). The heat release is proportional to  $t^{-1}$ (broken
line). In contrast to  NbTiH$_{9\%}$  the additional contribution
at short time was not observed.}
\end{figure}
In contrast to both NbTi and NbTiD$_{9\%}$ an additional nearly
exponential contribution to the heat release was observed for
NbTiH$_{9\%}$ at temperatures  $T_1 > 8$ K. Notice that this
behavior is not a consequence of different cooling: the cooling rate
was the same for NbTi, NbTiD$_{9\%}$ and NbTiH9$_{9\%}$.
At longer time the heat release was found to be again strongly
proportional to  $t^{-1}$ in agreement with Eq.~(\ref{eq6}). The
further analysis was performed just as described above for  NbTi.
Again, two kinds of TLSs with different freezing temperatures were
found. The separated contributions of both normal and anomalous
TLSs as well as the measured values of the heat release for $t_0 =
30$ min as a function of  $T_1^2-T_0^2$ are shown in Figs. 9 and
10.
\begin{figure} \centering
\includegraphics[width=8cm,  angle=-90]{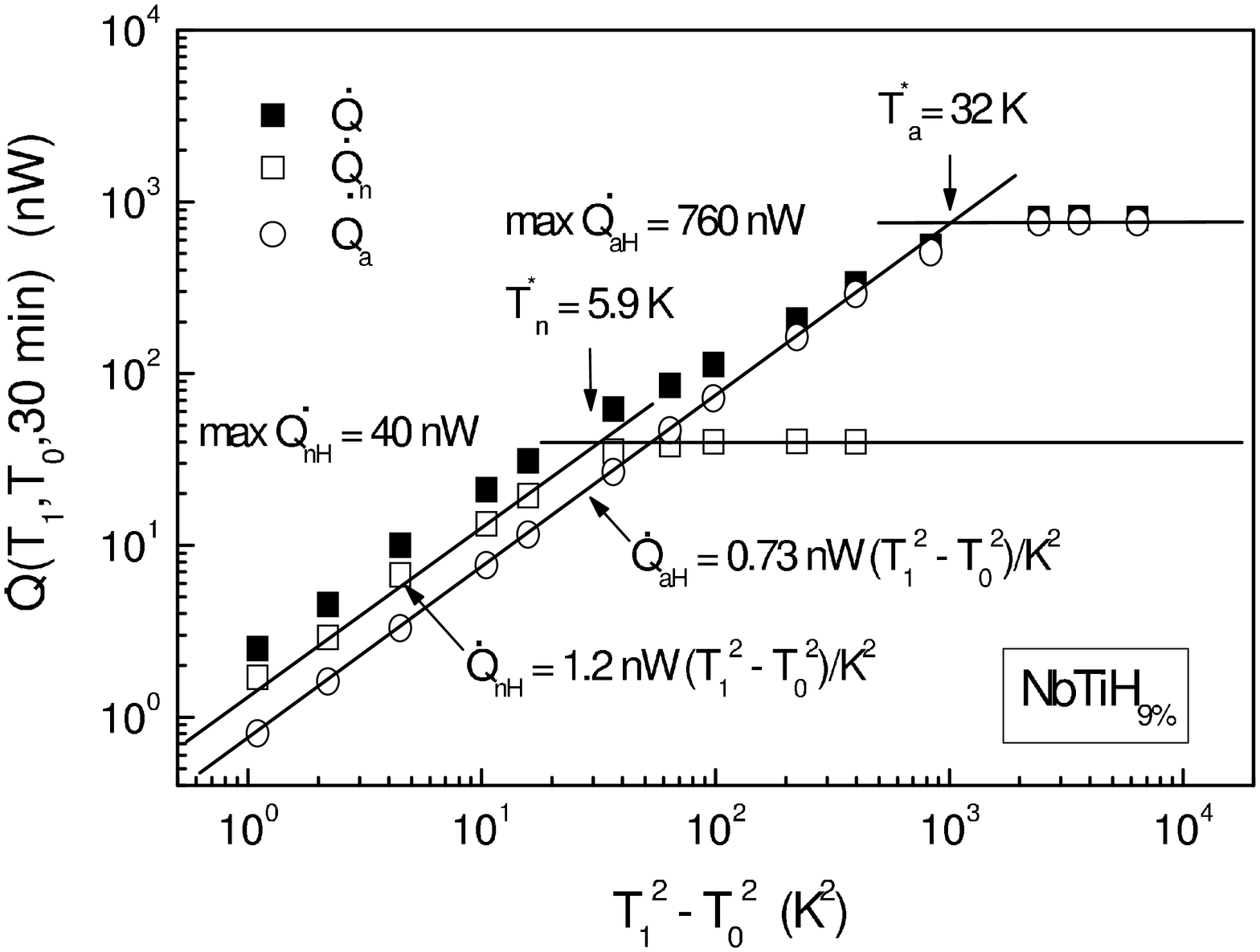}
\caption{The separated contributions of two kinds of TLSs to
the heat release in NbTiH$_{9\%}$   at $t_0 = 30$ min.  The white
squares and circles show the contribution of TLSs with the freezing
temperature  $T^*_n = 5.9$ K and  $T^*_a = 32$ K, correspondingly.
The sum of these two contributions yields the measured heat
release (black squares).}
\end{figure}
\begin{figure} \centering
\includegraphics[width=8cm,  angle=-90]{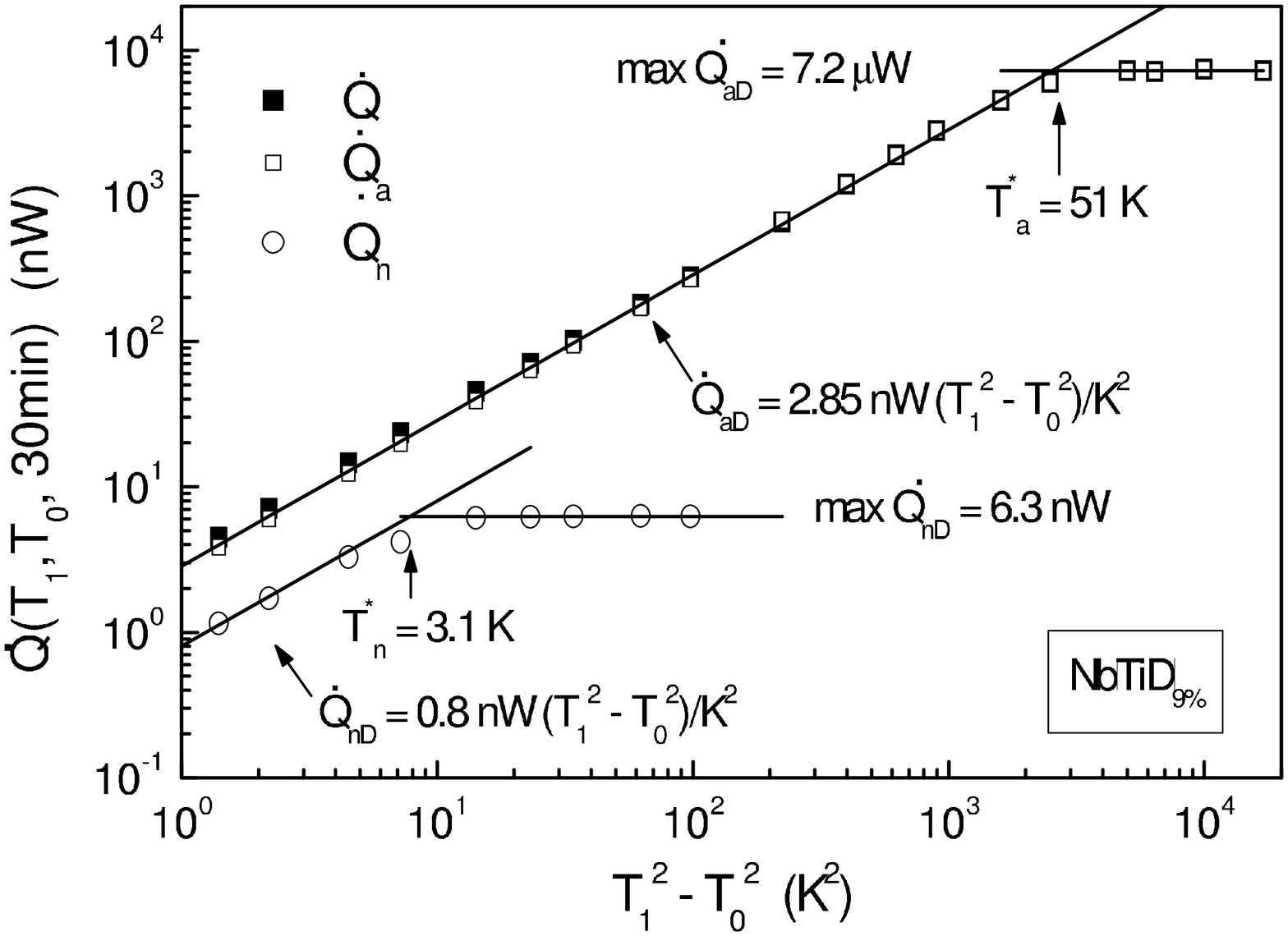}
\caption{The separated contributions of two kinds of TLSs to
the heat release in NbTiD$_{9\%}$   at $t_0 = 30$ min.  The white
circles and squares show the contribution of TLSs with the freezing
temperature  $T^*_n = 3.1$ K and  $T^*_a = 51$ K, correspondingly.
The sum of these two contributions yields the measured heat
release (black squares).}
\end{figure}
The corresponding parameters  $P_{Q}, T^*, V^{eff}, \Delta_{0},
\lambda$, and $E_{0}$ are given in Table II. Figs. 9 and 10 show
clearly the markedly different behavior of the isotopic effect for
the normal and anomalous TLSs. While the change of H to D leads to a
reduction of the maximum heat release of normal TLSs, the maximum
heat release of anomalous TLSs increases significantly. This
increase is mainly caused by an increase of $P_a$: $P_a^D/P_a^H =
3.8$(!)

Notice that in the soft potential model (SPM) the distribution
parameter $P$ is not a constant like in the STM and decreases weakly
with increasing $\lambda$ (see Ref.~\cite{Parshin2}). In this case
the absolute value of the heat release will be smaller than expected
from the STM. For instance in vitreous silica this gives roughly a
factor of 2 between the distribution parameter deduced from the heat
capacity and the heat release in better agreement with the
experimental results. In this case a different time dependence
should be observed for the heat release, i.e. the relaxation is
faster than $t^{-1}$. Numerical calculation within the SPM yields
for vitreous silica $t^{-1.1}$. However, the very precise heat
release data of vitreous silica give exactly $t^{-1}$, in agreement
with the STM (see Ref.~\cite{Sahling1}). Such a faster relaxation
was not also found for all other amorphous and glass like
crystalline solids.

At the same time, the saturation of the heat release above the
freezing temperature $T^*$ is exactly the same, since actually we
introduce this temperature from the SPM (see
Ref.~\cite{Parshin2}). Thus, the heat release data of Ca
stabilized ZrO$_2$ and NbTi-H, the unexpected isotopic effect
NbTi-H/D  and the discrepancies between the distribution
parameters deduced from the heat capacity and heat release cannot
be explained within the SPM.

\subsection{Isotopic effect}

For the normal contribution to the heat release we found the
expected isotopic effect (see Table III).
\begin{table}[th]\caption{~}
\begin{tabular}{||c||c|c||}
\hline
Parameter   &    experiment  &  calculated\\
\hline\hline
$P_{QnH}/P_{QnD}$            &1.47   &1.41\\
$T^*_{nH}/T^*_{nD}$          &1.90   &2.0\\
$V_{nH}^{eff}/V_{nD}^{eff}$              &1.97   &2.0\\
$\dot{Q}_{nH}^{max}/\dot{Q}_{nD}^{max}$  &6.3    &6.6\\
$E_{0n}^{H}/E_{0n}^{D}$      &1.93   &2.0\\
$P_C^H/P_C^D$                &1.40   &1.41\\
$P_{Qa}^H/P_{Qa}^D$          &0.26   &1.41\\
$T^{*H}_{a}/T^{*D}_{a}$          &0.63   &2.0\\
$\dot{Q}_{aH}^{max}/\dot{Q}_{aD}^{max}$  &0.11   &6.6\\
$E_{0a}^{H}/E_{0a}^{D}$      &0.66   &2.0\\
\hline
\end{tabular}
\end{table}
We also observed the calculated isotopic effect for the distribution
parameters deduced from the heat capacity  $P_C^H/P_C^D = 1.40$
(see Tables II and III). At the same time, for the anomalous TLSs
we found a rather unexpected behavior of the distribution parameters,
freezing temperatures and the maximum contribution to the heat release (see
Table III and the discussion below).

\subsubsection{Comparison between the data of the heat capacity and the heat
release}

The characteristic feature of anomalous TLSs is their surprisingly
high freezing temperature and, correspondingly, the large barrier
height (at least at the freezing temperature). In amorphous
dielectrics an upper limit of the barrier height distribution was
found to lie in the range of 600 K (a - SiO$_2$ \cite{Sahling1,
Keil}) and 3130 K (a - B$_2$O$_3$ \cite{Rau}). Thus, the barrier
heights of anomalous TLSs are of the order of the maximum barrier
height. After cooling, the corresponding maximum tunneling
parameter of normal TLSs is $\lambda_n^{max} = V_n^{max}/E_{0n}$
and of the anomalous ones $\lambda_{a}^{max}$, which is
essentially smaller $\lambda_n^{max}$. Let us denote the effective
tunneling parameter obtained in the heat release measurements as
$\lambda_Q$. Generally, one has to consider three possible cases:
$\lambda_a^{max} < \lambda_Q$, $\lambda_a^{max} > \lambda_Q$ and
$\lambda_a^{max} = \lambda_Q$. In the first case, the anomalous
TLSs will not contribute to the heat release, however, they will
increase the heat capacity, i.e. $P_C > P_Q$. This was observed in
NbTi, NbTi-H and a-SiO$_2$ (see Table IV). For $\lambda_a^{max} >
\lambda_Q$, the anomalous TLSs contribute to both the heat
capacity and the heat release and we expect that $P_C = P_Q$. In
addition, we have observed a giant heat release after cooling from
$T_1 > T^*_a$ originated from the high freezing temperature of
anomalous TLSs. This is the case for NbTi-D, where $P_C/P_Q = 1$.
The most interesting case is $\lambda_Q = \lambda_a^{max}$ where
we expect to reveal an exponential decrease of the heat release
with time at the corresponding maximum relaxation time
$\tau_a^{max}=\tau(V^{max}_a)$. Indeed, such behavior was observed
in (ZrO$_2$)$_{0.89}$(CaO)$_{0.11}$ \cite{Sahling2} (see Fig.~11).
\begin{figure} \centering
\includegraphics[width=8cm,  angle=-90]{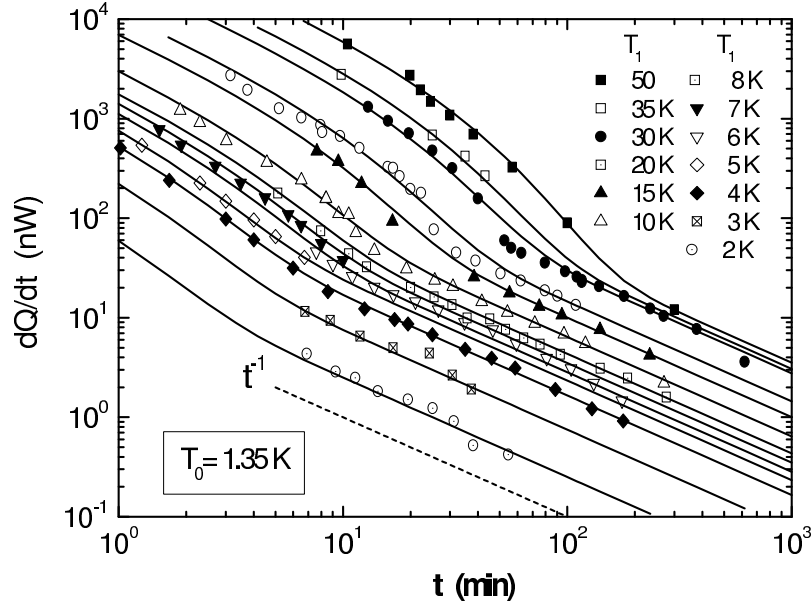}
\caption{The heat release in 14.5 cm$^3$
(ZrO$_2$)$_{0.89}$(CaO)$_{0.11}$ after cooling from different
initial temperatures  $T_1$  to the final phonon temperature $T_0$
as a function of time ($t = 0$ at the beginning of cooling)
\cite{Sahling2}. The curves are calculated with Eq.~(\ref{eq15}),
which corresponds to the distribution $P(\tau)$ of the standard
tunneling model with a step at $\tau_a^{max}$ caused by the
cut-off in the distribution of anomalous TLSs. The fit parameters
$Q_s = Q_{as} + Q_n$ , $Q_l = Q_{al} + Q_{n}$ and $\tau_a^{max}$
are given in Figs.~13 and 14. }
\end{figure}
A good fit of these heat release data was obtained with
\begin{equation}\label{eq15}
\dot{Q} =  Q_s t^{-1} \exp(-t/\tau_a^{max}) + Q_l t^{-1}.
\end{equation}
At $t < \tau_a^{max}$ the giant heat release was observed and $P_C =
P_Q$ (see Table IV).
\begin{table}[th]\caption{~}
\begin{tabular}{||c||c|c|c|c|c||}
\hline
Material      &$P_C$ ($10^{44}$ J$^{-1}$m$^{-3}$) &$P_Q$ ($10^{44}$ J$^{-1}$m$^{-3}$) &$P_C/P_Q$ &case         &Ref.\\
\hline\hline
a - SiO$_2$   &8.0                                &2.0                                 &4.0      &$\lambda_a^{max} < \lambda_Q$ & \cite{Sahling1}\\
NbTi          &43.2                               &5.3                                 &8.2      &$\lambda_a^{max} < \lambda_Q$ & ~\\
NbTiH$_{9\%}$ &120                                &45                                  &2.7      &$\lambda_a^{max} < \lambda_Q$ & ~\\
(ZrO$_2$)(CaO)&18.5                               &5.7                                 &3.2      &$\lambda_a^{max} < \lambda_Q$ & \cite{Sahling2}\\
(ZrO$_2$)(CaO)&18.5                               &19.2                                &0.96     &$\lambda_a^{max} > \lambda_Q$ & \cite{Sahling2}\\
NbTiD$_{9\%}$ &86                                 &84                                  &1.02     &$\lambda_a^{max} > \lambda_Q$ & ~ \\
\hline
\end{tabular}
\end{table}
For $t > \tau_a^{max}$  the contribution of anomalous TLSs reduces
significantly, but still remains nonzero. Thus, with the
distribution functions $P_a(\lambda)$ and $P_n(\lambda)$ shown in
Fig.~12 we can explain all results of the heat release and the
heat capacity measurements presented in Tables III and IV.
\begin{figure} \centering
\includegraphics[width=8cm,  angle=-90]{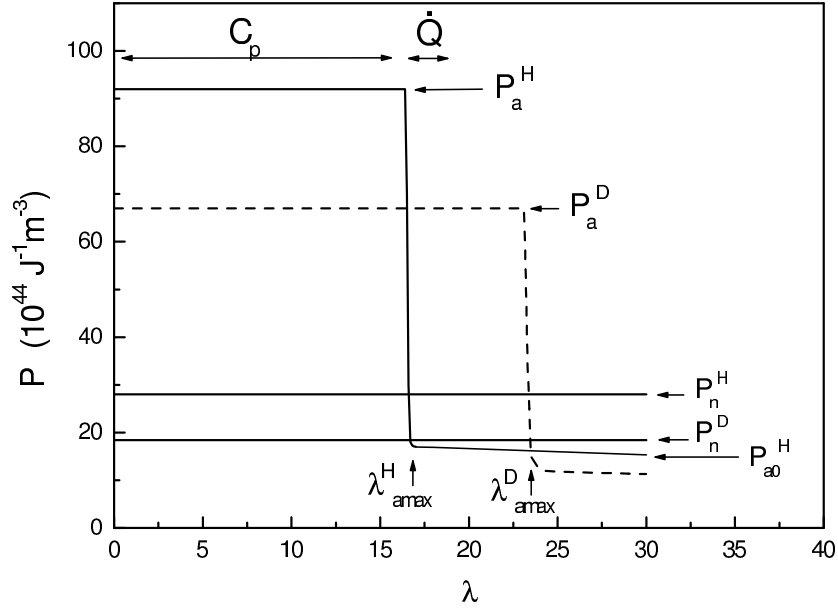}
\caption{The distribution functions of normal and anomalous TLSs as
a function of the tunneling parameter  $\lambda$   deduced from the
heat capacity and the heat release data of NbTi-H and NbTi-D. The
regions available from the heat capacity ($C_p$) and the heat
release ($\dot{Q}$) measurements are shown.}
\end{figure}

Important questions arise: what is the origin of $P_{a0}$  in
Fig.~12 and why did we observe a reduced but nevertheless a well
defined contribution of anomalous TLSs to the heat release for  $t
> \tau_a^{max}$? Moreover, comparing the corresponding barrier
heights with $V^{max}$ deduced from the acoustic experiments we
found that their values are always smaller. For example, in
vitreous silica $V_a^{eff}/k_B = 520$ K while $V^{max}/k_B = 600$
K. In (ZrO$_2$)$_{0.89}$(CaO)$_{0.11}$ we found for $t <
\tau_a^{max}$ that $V_a^{eff}/k_B > 1700$ K in good agreement with
$V^{max}/k_B = 1800$ K deduced from the measured on the same
sample damping peak of internal friction \cite{Abens} while for $t
> \tau_a^{max}$ we got $V_a^{eff}/k_B = 1300$ K. This is also
valid for NbTi-H/D. In NbTi-H $\tau_a^{max}$ is shorter $t$ since
$P_C
> P_Q$, and the observed contribution to the heat release is caused
by $P_{a0}$. Their barrier height $V_a/k_B = 1300$ K. For NbTi-D,
$\tau_a^{max}$ is larger $t$ and the barrier height of anomalous
TLSs causing the giant heat release is  $V_a^{eff}/k_B = 1750$ K
($t_0 = 30$ min).

In order to answer these questions one has to assume the existence
of some process leading to a drastic reduction of $\lambda_{max}$ to
$\lambda_a^{max}$ after cooling the sample. For example, if the
origin of such process is the mechanical deformation of the sample
during the rapid cooling, we would have got some distribution of the
internal stresses in the sample. In this case, the normal TLSs will
be situated in the regions with a minimal stresses. A maximum change
of $\lambda_a$ happens to be in the regions with the maximum
deformations. In the intermediated regions there will exist TLSs
with $\lambda$-values between $\lambda_{max}$ and $\lambda_a^{max}$,
which could cause $P_{a0}$ in the distribution function
$P_a(\lambda)$. In principle, every process which transforms a part
of normal tunneling systems into anomalous ones will produce TLSs
with tunneling parameters between these two groups. The appearance
of a contribution to the heat release at $t> \tau_a^{max}$ and
$\lambda_Q> \lambda_a^{max}$ is probably the consequence of this
fact. From this point of view it is not surprising that for all
amorphous and glass-like crystalline solids, where $P_C > P_Q$, a
contribution of the anomalous TLSs to the heat release was observed.

\subsubsection{Possible nature of anomalous TLSs}

As it follows from measured $T^*$ anomalous systems should have
large barrier heights. On the other hand they are able to relax in
experiment time and give contributions to the heat release and the
heat capacity. Furthemore, in accordance with Eq.~(\ref{eq15}) the
edge of the distribution function $P_a(\lambda)$ can be extracted
from the heat release measurements in both NbTi-H and
(ZrO$_2$)$_{0.89}$(CaO)$_{0.11}$. $\lambda^{max}_a$ can be shifted
more to the left (like for NbTi and a-SiO$_2$) or more to the
right (NbTi-D) (see Fig.~12). This is also clearly seen from the
wrong isotopic relations for anomalous TLSs (see Table III) as
well as from the relation $P_C/P_Q$ (see Table IV).

As was mentioned above this situation could be explained by either
markedly different values of $E_{0n}$ and $E_{0a}$ or some kind of
instability which influences barrier heights. Since the correct
isotopic relation is not fulfilled for observed $E_{0a}$, it is
naturally to suggest that $E_{0a}=E_{0n}=E_{0}$. Therefore, the
second possibility looks more reasonable. Indeed, in
Ref.~\cite{Karpov} a generation of strong local mechanical
stresses $\sigma$ during the cooling of sample was suggested. In
this case, the fluctuations in thermal expansion of glass-like
system can reach giant values. According to \cite{Karpov}, the
thermal expansion coefficients $\alpha(\textbf{r})$ take random
values inside the dilatation centers of the sample. The dispersion
in the distribution of $\alpha(\textbf{r})$ is suggested to be
much bigger of its mean value: $\langle\alpha^2\rangle \gg
\langle\alpha\rangle^2$. On the other hand,
\begin{equation}\label{disp}
\frac{\langle\alpha\rangle^2}{\langle\alpha^2\rangle} \sim
\frac{\Gamma^2}{\gamma^2} \sim \frac{\Gamma^2 (k_BT)^2}{D^2},
\end{equation}
where $\Gamma$ and $\gamma$ are the 'global' ('local') Gruneisen
parameters, respectively, and $D$ is the deformation potential.
Supposing the normal (Gaussian) distribution of the dilatation
centers, the maximum value of $\alpha T$ is estimated as
\begin{equation}\label{eq22}
(\alpha T)^{max} \simeq \sqrt{2 \langle\alpha^2\rangle}T \sim
\frac{\sqrt{2} \langle\alpha\rangle D}{\Gamma k_B}.
\end{equation}

The stresses at the local thermal expansion are written as
\begin{equation}
|\sigma| = K \alpha \Delta T,
\end{equation}
where $K$ is the shear modulus. In accordance with the Eyring model
an applied stress results in a linear reduction of the barrier
height \cite{Eyring}
\begin{equation}
\Delta V = \pm \sigma V_{ac},
\end{equation}
where $V_{ac}$ is a so-called activation volume (a typical volume
required for a molecular shear rearrangement). Let $\Delta
V^{min}$ be the necessary reduction of the barrier height to make
the relaxation time comparable with the experiment time, i.e.
$\tau(V^{max}_a) = t_0$, where $V^{max}_a=V^{max}_n - \Delta
V^{min}$. Then the minimum value of $\alpha\Delta T$ reads
\begin{equation}
(\alpha \Delta T)^{min} = \frac{\Delta V^{min}}{K V_{ac}}.
\end{equation}
Therefore, the fulfilment of the condition $(\alpha
T)^{max}\geq(\alpha \Delta T)^{min}$ would be a signal that
anomalous TLSs can be produced by random local stresses. Taking
typical values $\Gamma\sim 30$, $\Delta V^{min}/k_B = 1.5\times
10^3$ K, $\langle\alpha\rangle\sim 10^{-8}$ K$^{-1},$ $D\sim 1$
eV, $K\sim 10^{11}$ Pa, $V_{ac}\sim 10$ nm$^{3}$, one obtains that
$(\alpha T)^{max}\sim (\alpha \Delta T)^{min}$. This estimation
allows us to conclude that mechanical deformations can generate
anomalous TLSs.

By fitting with Eq.~(\ref{eq15}) the heat release data for
(ZrO$_2$)$_{0.89}$(CaO)$_{0.11}$ and NbTiH$_{9\%}$  we can deduce
also the dependence of $\tau_a^{max}$ on the average energy
$E_{av}/k_B =  T_1 + T_0$ (see Fig.~14).
\begin{figure} \centering
\includegraphics[width=8cm,  angle=-90]{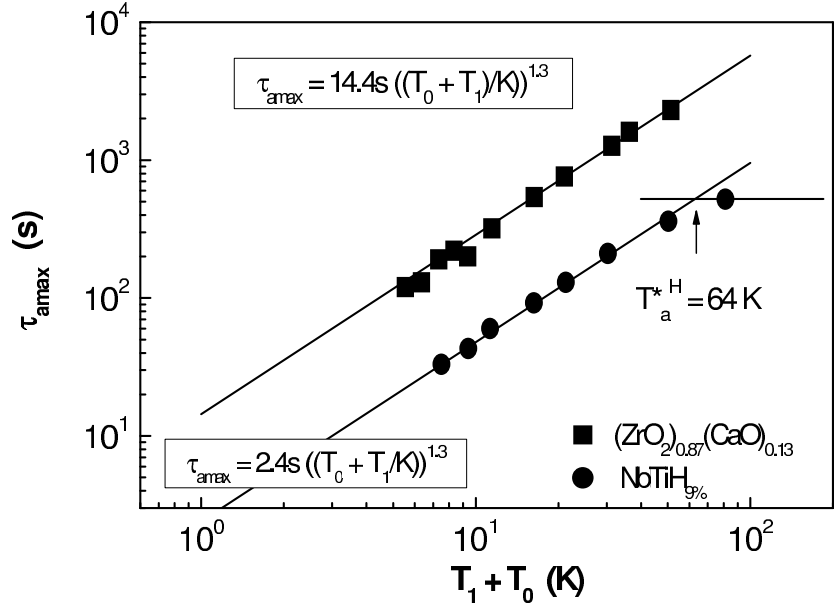}
\caption{The fit parameters  $\tau_a^{max}$  used to fit curves
for the heat release data of (ZrO$_2$)$_{0.89}$(CaO)$_{0.11}$ (see
Fig.~11) and   NbTiH$_{9\%}$ (see Fig.~7) as a function of the
average energy of TLSs  $E_{av}/k_B = T_1 + T_0$. The maximum
relaxation time is proportional to $E^{1.3}$ for both materials.
The saturation of $\tau_a^{max}$ for NbTiH$_{9\%}$ at high $T_1$
allows one to estimate the freezing temperature  $T^{H*} = 64$ K.}
\end{figure}
As is seen from Fig.~14,  $\tau_a^{max}$ is roughly proportional
to $T_1 + T_0$. This can be explained by the fact that the sample
was repeatedly cooled and warmed during experiment so that every
next cycle is accompanied by a small relaxation of internal
stresses (there is a hysteresis).

\subsubsection{Explanation of Isotopic effect of anomalous tunneling systems\label{435}}

The assumption that the cut-off in the distribution function of
anomalous TLSs is caused by the maximum barrier height of normal
TLSs  $V^{max}_n$ with $E_{0a}=E_{0n}=E_0$, i.e.
\begin{equation}\label{eq16}
\lambda_a^{max} \sim  \sqrt{m V^{max}_a}.
\end{equation}
also explains the observed isotopic effect. Taking into account
that $V^{max}_n$ and $\Delta V$ are defined by NbTi matrix we
conclude that $V^{max}_{aH} = V^{max}_{aD}$ being independent from
the mass of TLSs. Therefore the relation between anomalous
tunneling parameters takes the following form:
\begin{equation}\label{eq17}
\frac{\lambda_a^{D max}}{\lambda_a^{H max}} = \sqrt{\frac{m^D}{m^H}}
= \sqrt{2}.
\end{equation}

Since  $P_C = P_a + P_n$ and $P_a\gg P_n$, we expect to obtain
$P_C^H/P_C^D \approx \sqrt{2}$ and this relation was indeed
observed (see Table III). Thus, according to Eq.~(\ref{eq17}) with
an increase of $\lambda_a^{max}$ the case $\lambda_a^{max} <
\lambda_Q$ (NbTi-H) is changed by $\lambda_a^{max} > \lambda_Q$
(NbTi-D). As discussed above, in the first case we have $P_C = P_a
+ P_n$, $P_Q = P_{a0} + P_n$ with $P_{a0}\ll P_a$ and, hence,
$P_C/P_Q
>1$. In the second case, one has $P_C = P_Q = P_a + P_n$, $P_C/P_Q = 1$ and we
observe a giant heat release. All these conclusions agree with the
experimental results. Moreover, we can explain now also the
additional exponential contribution to the heat release observed
just after the cooling of NbTi-H. This behavior is seen when
$\tau_a^{H max}$ is close to the time necessary for cooling the
sample. Let us analyze the heat release data of NbTi-H with
Eq.~(\ref{eq15}) where $Q_s$ is determined by $P_C$,
\begin{equation}\label{eq18}
Q_s = \frac{\pi^2k_B^2}{24} P_C V (T_1^2 - T_0^2),
\end{equation}
and  $Q_l =  t\dot{Q}$ for  $t >  \tau_a^{max}$ with the only free
fitting parameter $\tau_a^{max}$.

This fit describes the experimental data shown in Fig.~12
perfectly well with an energy dependence of $\tau_a^{max}$ similar
to that observed for (ZrO$_{2}$)$_{0.89}$(CaO)$_{0.11}$ (see
Fig.~14).
\begin{figure} \centering
\includegraphics[width=8cm,  angle=-90]{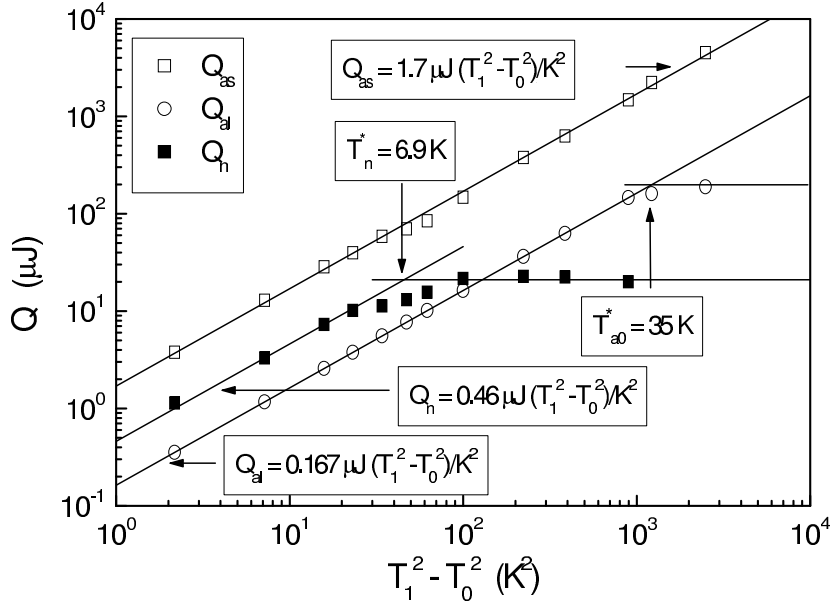}
\caption{$Q_n$,  $Q_{as}$ and  $Q_{al}$ as a function of $T_1^2 -
T_0^2$. The analysis of the fit parameters  $Q_s$  and $Q_l$ used
to fit curves in Fig.~11 shows that at short time $Q_s = Q_{as} +
Q_{n}$ and at long time $Q_l = Q_{al} + Q_{n}$.  $Q_{as}$ and
$Q_{al}$ determine the contribution of anomalous TLSss for short
($t \ll  \tau_a^{max}$) and long time ($t \gg  \tau_a^{max}$) and
differ by a factor of 10. Together with the reduction of the heat
release caused by the anomalous TLSss we observe a decrease of
their freezing temperature: for $Q_{as}$ one has $T^*_a > 50$ K,
while $T^*_a=35$ K for $Q_{al}$. The contribution of normal TLSss
$(Q_n)$ at long and short time is unchanged. }
\end{figure}
Thus, $\tau_a^{H max}  = 10^2$ s with a corresponding $\lambda_a^{H
max} = 17$. The estimation of $\lambda_a^{D max}$ with
Eq.~(\ref{eq17}) gives $\lambda_a^{D max}=24$ with a corresponding
$\tau_a^{D max} = 10^8$ s, which is much larger our longest
measuring time of $10^5$ s in the heat release experiment with
NbTi-D.

Notice that the effective masses can be influenced by the dressing
effect (see, e.g., \cite{Sethna}). However, in our case this
effect cannot give any reasonable explanation of the results,
since it is necessary that the effective mass of H must be
increased by a factor of 4, while at the same time the effective
mass of D remains unchanged. In addition, the coupling constants
between the tunnelling systems and the lattice are very small for
H and D (see Table I).

\section{Conclusion}

In all investigated materials, NbTi, NbTi-H, and NbTi-D, the
low-temperature anomalies typical for structural glasses have been
observed for main investigated characteristics: the thermal
conductivity, the heat capacity and the heat release.

The thermal conductivity of NbTi-H/D increases roughly 3 times for
concentrations of H or D larger a critical concentration of 2\%. H
or D with concentrations larger 2\% stabilize the $\beta$-phases
of NbTi. In this way, the TLSs of the NbTi matrix which are caused
by local fluctuations between the $\beta$-  and $\omega$- phases
disappear. Thus, at concentrations higher than the critical one,
the glassy behavior of NbTi-H/D is determined by the TLSs of H or
D only and we have a good system to investigate the isotopic
effect of TLSs in glasses.

For the heat capacity we found the expected isotopic effect with the
distribution parameter  $P_C^H$ being about 40\% larger the value
$P_C^D$.

Two kinds of TLSs contribute to the heat release  at a given time.
This follows from their different freezing temperatures.
Consequently, TLSs with very different barrier heights exhibit the
same relaxation time and contribute at the same time to the heat
release. This result agrees with that observed in all other
investigated amorphous and glass-like crystalline materials.

For TLSs with the lower barrier heights we have observed all
isotopic effects estimated within the standard tunneling model.
Namely, the distribution parameter  $P_{Qn}^H$ is about 40\% larger
$P_{Qn}^D$, the freezing temperature  $T^{*H}$ is roughly 2 times
larger the value $T^{*D}$, while the ratio of maximum values of the
heat release gives the expected factor close to 6.

A completely different behavior we have discovered for the TLSs with
the higher freezing temperatures, which we call anomalous TLSs.
Instead of the expected reduction of the distribution parameter
$P_Q^D$ in comparison with $P_Q^H$ of about 40\% we have observed
its increase with the factor of 3.8. The freezing temperature
$T^{D*}_{a}$ is found to be larger the freezing temperature
$T^{H*}_{a}$ instead of the expected reduction by the factor of 2
and we have observed a giant maximum value of the heat release,
roughly 10 times larger the maximum value in NbTi-H, in contrast to
the 6 times smaller expected value. In addition, the distribution
parameter $P_C^H$ is much larger the corresponding value deduced
from the heat release data $P_Q^H$ $(P_C^H/P_Q^D = 2.7)$, while for
NbTi-D we found an excellent agreement between these two parameters.
All these anomalous isotopic effects can not be explained within the
standard tunneling model.

We have suggested a simple explanation of these surprising results
by making an assumption that a cut-off in the distribution
function of anomalous TLSs exists. It can be caused by the cut-off
in the distribution of barrier heights at $V_a^{max}$ with the
corresponding maximum distribution parameter  $\lambda_a^{max}$.
We have noted three possibilities for the average value
$\lambda_Q$ of the TLSs contributing to the heat release:
$\lambda_a^{max} < \lambda_Q$,  $\lambda_a^{max} > \lambda_Q$ and
$\lambda_a^{max} = \lambda_Q$. In the first case, the anomalous
TLSs contribute to the heat capacity, but not to the heat release
(except for a small contribution which appears also for $\lambda >
\lambda_a^{max}$ with large barrier heights, yet smaller than
$V^{max}_n$). In this case, we have $P_C/P_Q > 1$ observed for
most of investigated amorphous and glass-like crystalline solids
including  NbTi and NbTi-H. In the second case, the anomalous TLSs
contribute to both the heat capacity and the heat release and
$P_C/P_Q = 1$. A characteristic feature of this case is a giant
heat release due to the high freezing temperature of the anomalous
TLSs. This is the case for NbTi-D. Thus, the anomalous isotopic
effect of the tunneling states in NbTi-H/D is caused by an
increase of $\lambda_a^{max}$ for NbTi-D in comparison with NbTi-H
due to the larger mass of the tunneling atoms at the same
distances and maximum barrier heights. Finally, in the most
interesting case $\lambda_a^{max} =  \lambda_Q$  the exponential
time dependence is expected. In fact, all three possible cases
were observed in (ZrO$_2$)$_{0.89}$(CaO)$_{0.11}$. This is a clear
experimental argument that our assumption on the existence of
anomalous TLSs with high freezing temperatures and, at the same
time, small tunneling parameters is quite reasonable. From
theoretical point of view, giant large-scale fluctuations in
thermal expansion are expected in glasses and related materials.
Our estimations show that arising local mechanical deformations
could produce anomalous TLSs.

\acknowledgements We thank P. Monceau and J.C. Lasjaunias (Neel
Institute, France) for the support of  the experimental work and
very helpful discussions. This work has been supported by the
Heisenberg-Landau program under grant No. HLP-2010-29.

\end{document}